\documentclass[11pt]{article}

\usepackage{amsmath}
\usepackage{graphicx}
\usepackage{amsfonts}
\usepackage{amssymb}
\usepackage{epsfig}
\usepackage{color}
\usepackage{psfrag}
\usepackage{epstopdf}

\usepackage{caption}
\usepackage{subcaption}

\newcommand{\EQ}{\begin{equation}}
\newcommand{\EN}{\end{equation}}
\newcommand{\be}{\begin{equation}}
\newcommand{\ee}{\end{equation}}
\newcommand{\bea}{\begin{eqnarray}}
\newcommand{\eea}{\end{eqnarray}}

\begin{document} \setcounter{page}{0}
\newpage
\setcounter{page}{0}
\renewcommand{\thefootnote}{\arabic{footnote}}
\newpage
\begin{titlepage}
\begin{flushright}
%SISSA 40/2012/EP \\
%DFTT 9/2007
\end{flushright}
\vspace{0.5cm}
\begin{center}
{\large {\bf On unitary time evolution out of equilibrium}}\\
\vspace{1.8cm}
{\large Gesualdo Delfino$^{1,2}$ and Marianna Sorba$^{1,2}$}\\
\vspace{0.5cm}
{\em $^1$SISSA -- Via Bonomea 265, 34136 Trieste, Italy}\\
{\em $^2$INFN sezione di Trieste, 34100 Trieste, Italy}\\
%{\em E-mail: delfino@sissa.it}\\
%\vspace{0.5cm}
%{\large and}\\
%\vspace{0.5cm}
%{\large P. Simonetti}\\
%\vspace{0.5cm}
%{\em Department of Physics, University of Wales Swansea,\\
%Singleton Park, Swansea SA2 8PP, United Kingdom}\\
%{\em email: p.simonetti@swansea.ac.uk}\\
\end{center}
\vspace{1.2cm}

\renewcommand{\thefootnote}{\arabic{footnote}}
\setcounter{footnote}{0}

\begin{abstract}
\noindent
We consider $d$-dimensional quantum systems which for positive times evolve with a time-independent Hamiltonian in a nonequilibrium state that we keep generic in order to account for arbitrary evolution at negative times. We show how the one-point functions of local operators depend on the coefficients of the expansion of the nonequilibrium state on the basis of energy eigenstates. We express in this way the asymptotic offset and show under which conditions oscillations around this value stay undamped at large times. We also show how, in the case of small quenches, the structure of the general results simplifies and reproduces that known perturbatively.
\end{abstract}
\end{titlepage}

\newpage
\tableofcontents

\section{Introduction}
The problem of the fate at large times of an extended and isolated quantum system out of equilibrium was addressed analytically in \cite{BMcD} for the case of noninteracting fermions in one dimension. Understanding the case of interacting quasiparticles turned out to be a difficult problem that could be faced through the perturbative approach introduced in \cite{quench} and further developed in \cite{DV,oscill,oscillD,excited}. The perturbative analysis applies to the basic way of dynamically generating nonequilibrium evolution, which has been called "quantum quench" \cite{SPS,CC} in analogy with thermal quenches of classical statistical systems: the system is in a stationary state until the sudden change of an interaction parameter leads to the new Hamiltonian that rules the unitary time evolution thereafter. The theory is perturbative in the quench size (see Eq.~(\ref{quench})), while the interaction among the quasiparticles can be arbitrarily strong. It allows to actually {\it determine} the nonequilibrium state generated by the quench and to follow analytically the time evolution. One result is that, in any spatial dimension, interaction leads to persistent oscillations of one-point functions (e.g. the order parameter) when the state produced by a homogeneous quench includes a one-quasiparticle mode and no internal symmetry prevents the observable to couple to that mode \cite{quench}. It was observed in \cite{oscill} that, if undamped oscillations are present at first order in the quench size, they will remain as a feature of the nonperturbative result, a prediction that found a remarkable confirmation in \cite{Jacopo}, where no decay of the oscillations was observed in a simulation of the Ising chain reaching times several orders of magnitude larger than the perturbative timescale. The undamped oscillations of \cite{quench} have also been observed in simulations performed over shorter time intervals, see \cite{BCH,KCTC,Lukin,Liu,EK} for a nonexhaustive list and \cite{oscill,oscillD} for a discussion of the different instances. 

It is the purpose of this paper to further investigate the properties of unitary nonequilibrium evolution at large times. To this end, we consider a quantum system in $d$ spatial dimensions with Hamiltonian
\EQ
\left\{
\begin{array}{l}
H_0({\bf x},t)\,,\hspace{.5cm}t<0\,,\\
\\
H,\hspace{1.5cm}t\geq 0\,,
\end{array}
\right.
\label{Hamiltonian}
\EN
where $H$ does not depend on time and is also translation invariant in space. We are interested in the time evolution of the system for $t>0$, and expand the nonequilibrium state on the basis of the quasiparticle states $|{\bf p}_1,\ldots,{\bf p}_n\rangle$ of the Hamiltonian $H$, with coefficient functions $f_n({\bf p}_1,\ldots,{\bf p}_n)$ which know about the evolution of the system since $t=-\infty$. The main question we want to answer is how the one-point function of a local operator $\Phi$ depends on the coefficients $f_n$. Hence, in order to study the features of the generic case, these coefficients are not specified and we perform a nonperturbative analysis relying on the structural properties of unitary time evolution of quasiparticle modes. While we analyze the problem in a more general way, we anticipate here the result for the fully translation invariant case (no spatial inhomogeneities are inherited from $t<0$) and large times; we obtain
\begin{align}
\langle \Phi(t)\rangle =A_\Phi+\frac{F_{1}^{\Phi}}{(2\pi)^d M}(h_1\,e^{-iMt}+h^*_1\,e^{iMt})+ O(t^{-d/2})\,,\hspace{1cm}t\to\infty\,,
\label{asympt}
\end{align}
where $F_1^\Phi$ is the one-quasiparticle form factor of the operator $\Phi$ and $M$ is the quasiparticle mass. We show that all the $f_n$'s enter the determination of $h_1$ and of the asymptotic offset $A_\Phi$, in a way that we find out. On the other hand, for dynamically generated nonequilibrium evolution, the condition $h_{1}F_1^\Phi\neq 0$ for the presence of the undamped oscillations in (\ref{asympt}) amounts to ${f}_0{f}_{1}F_1^\Phi\neq 0$ on symmetry grounds.

For small quenches, which are a particular case of (\ref{Hamiltonian}), we show how the structure of the present general results simplifies and reduces to that known from perturbation theory. While the result (\ref{asympt}) is written for the case of a single quasiparticle species, we will also discuss the generalization to several species. 

The paper is organized as follows. In the next section we consider the problem of one-point functions in the general case of the Hamiltonian (\ref{Hamiltonian}), while in section~3 we analyze the fully translation invariant case. The way the general results reduce to those known perturbatively for small quenches is shown in section~4 before providing some final remarks in the last section. An appendix reviews some historical developments.

\section{General setting}
We consider a $d$-dimensional quantum system with the Hamiltonian (\ref{Hamiltonian}), where $H$ does not depend on time and is also translation invariant in space. We are interested in the time evolution for positive times, which we investigate considering the expectation values $\langle \Phi({\bf x},t)\rangle$ of local, scalar, Hermitian operators $\Phi({\bf x},t)$ on the quantum state of the system. We consider that at $t=-\infty$ the system is in the ground state $|\Omega\rangle$ of its Hamiltonian $H_0({\bf x},-\infty)$, with the normalization $\langle\Omega|\Omega\rangle=1$. Then, if $I$ denotes the identity operator, the condition
\EQ
\langle \Phi({\bf x},t)\rangle=1\hspace{.5cm}\textrm{if}\hspace{.5cm}\Phi=I
\label{normalization}
\EN
holds at $t=-\infty$ and is henceforth preserved by the time evolution.

We have
\begin{equation}
\Phi(\mathbf{x},t)=e^{i(\boldsymbol{\mathcal{P}}\cdot \mathbf{x}+Ht)}\Phi(0,0) e^{-i(\boldsymbol{\mathcal{P}}\cdot \mathbf{x}+Ht)}\,,\hspace{1cm}t\geq 0\,,
\label{shift}
\end{equation}
where $\boldsymbol{\mathcal{P}}$ denotes the momentum operator. The state $|\psi\rangle$ of the system can be generally expanded on the basis of asymptotic quasiparticle\footnote{Quasiparticles are collective excitation modes commonly exhibited by statistical systems. For instance, a very general case in which they arise is the proximity of critical points (e.g. in spin models or quantum gases).} states $|{\bf p}_1,\ldots,{\bf p}_n\rangle$ of the theory with Hamiltonian $H$, which are eigenstates of energy and momentum with eigenvalues 
\begin{equation}
E=\sum_{i=1}^n E_{\mathbf{p}_i}\,,\hspace{1cm} \mathbf{P}=\sum_{i=1}^n \mathbf{p}_i\,,
\end{equation}
respectively. Energy and momentum of the quasiparticles are related as 
\EQ
E_{\mathbf{p}}=\sqrt{M^2+\mathbf{p}^2}\,, 
\label{dp}
\EN
where $M>0$ is the quasiparticle mass and measures the distance from a quantum critical point\footnote{In general, statistical systems allow for critical points in their parameter space, and relativistic invariance emerges in their proximity. At the same time the formalism includes the nonrelativistic case as a low energy limit. For the sake of notational simplicity we refer to the case of a single quasiparticle species; generalizations will be discussed when relevant.}; we also adopt the state normalization
\begin{equation}
\langle \mathbf{q}|\mathbf{p}\rangle=(2\pi)^d E_{\mathbf{p}}\, \delta(\mathbf{q}-\mathbf{p})\,.
\end{equation}
The expansion of the state $|\psi\rangle$ on the basis of asymptotic quasiparticle states takes the form
\begin{equation}
|\psi\rangle = \sum_{n=0}^{\infty}\int_{-\infty}^{\infty} \prod_{i=1}^n \frac{d\mathbf{p}_i}{(2\pi)^d E_{\mathbf{p}_i}}\, f_n(\mathbf{p}_1,...,\mathbf{p}_n)\, |\mathbf{p}_1,...,\mathbf{p}_n\rangle\,,
\label{state}
\end{equation}
where the coefficient functions $f_n({\bf p}_1,\ldots,{\bf p}_n)$ give the probability amplitude that the state $|\mathbf{p}_1,...,\mathbf{p}_n\rangle$ is observed at $t=+\infty$. 

The one-point function $\langle\Phi(\mathbf{x},t)\rangle$ is continuous at $t=0$, although in general non-differentiable, and continuity at $t=0$ is ensured writing
\EQ
\langle\Phi(\mathbf{x},t)\rangle=G_\Phi(\mathbf{x},t)-G_\Phi(\mathbf{x},0)+\langle\Phi(\mathbf{x},0)\rangle\,,
\label{continuity}
\EN
where
\EQ
G_\Phi(\mathbf{x},t)=\langle \psi|\Phi(\mathbf{x},t)|\psi\rangle\,.
\label{G}
\EN
Recalling (\ref{state}), we have
\begin{align}
G_\Phi(\mathbf{x},t)&=\sum_{n,m=0}^{\infty}\int\prod_{i=1}^n \frac{d\mathbf{p}_i}{(2\pi)^d E_{\mathbf{p}_i}}\, \prod_{j=1}^m \frac{d\mathbf{q}_j}{(2\pi)^d E_{\mathbf{q}_j}}\,  f_n(\mathbf{p}_1,...,\mathbf{p}_n)f^*_m(\mathbf{q}_1,...,\mathbf{q}_m) \nonumber\\
&\times F_{m,n}^{\Phi}(\mathbf{q}_1,...,\mathbf{q}_m|\mathbf{p}_1,...,\mathbf{p}_n)\,e^{i[(\mathbf{Q}-\mathbf{P})\cdot \mathbf{x}+(\tilde{E}-E)t]}\,,\hspace{2cm}t\geq 0\,,
\label{G1}
\end{align}
where we used (\ref{shift}) and defined
\begin{equation}
\tilde{E}=\sum_{j=1}^m E_{\mathbf{q}_j}\,,\qquad \mathbf{Q}=\sum_{j=1}^m \mathbf{q}_j\,,
\end{equation}
and
\begin{equation}
F_{m,n}^{\Phi}(\mathbf{q}_1,...,\mathbf{q}_m|\mathbf{p}_1,...,\mathbf{p}_n)= \langle \mathbf{q}_1,...,\mathbf{q}_m|\Phi(0,0)|\mathbf{p}_1,...,\mathbf{p}_n\rangle\,.
\label{matrixelement}
\end{equation}
The matrix elements (\ref{matrixelement}) decompose into the sum of a connected term 
\begin{equation}
F^{\Phi,c}_{m,n}(\mathbf{q}_1,...,\mathbf{q}_m|\mathbf{p}_1,...,\mathbf{p}_n)=\langle \mathbf{q}_1,...,\mathbf{q}_m |\Phi(0,0)|\mathbf{p}_1,...,\mathbf{p}_n\rangle_c\,,
\label{connectedmatrixelement}
\end{equation}
plus disconnected terms containing delta functions associated to the annihilations of particles on the left with particles on the right, namely\footnote{We refer to bosonic statistics, which is generic for quasiparticles in $d>1$, and comment in section~\ref{pt} about the case $d=1$.}
\begin{align}
&F_{m,n}^{\Phi}(\mathbf{q}_1,...,\mathbf{q}_m|\mathbf{p}_1,...,\mathbf{p}_n)= F^{\Phi,c}_{m,n}(\mathbf{q}_1,...,\mathbf{q}_m|\mathbf{p}_1,...,\mathbf{p}_n)+\sum_{i=1}^n \sum_{j=1}^m (2\pi)^d E_{\mathbf{p}_i} \delta(\mathbf{p}_i-\mathbf{q}_j)\nonumber\\
&\times F^{\Phi,c}_{m-1,n-1}(\mathbf{q}_1,...,\mathbf{q}_{j-1},\mathbf{q}_{j+1},...,\mathbf{q}_m|\mathbf{p}_1,...,\mathbf{p}_{i-1},\mathbf{p}_{i+1},...,\mathbf{p}_n)\nonumber\\
&+ \sum\limits_{\substack{i,l=1\\i\neq l}}^n \sum\limits_{\substack{j,k=1\\j\neq k}}^m (2\pi)^{2d} E_{\mathbf{p}_i} E_{\mathbf{p}_l} \delta(\mathbf{p}_i-\mathbf{q}_j) \delta(\mathbf{p}_l-\mathbf{q}_k)\nonumber\\
&\times F^{\Phi,c}_{m-2,n-2}(\mathbf{q}_1,...,\mathbf{q}_{j-1},\mathbf{q}_{j+1},...,\mathbf{q}_{k-1}, \mathbf{q}_{k+1},...,\mathbf{q}_m|\mathbf{p}_1,...,\mathbf{p}_{i-1},\mathbf{p}_{i+1},...,\mathbf{p}_{l-1}, \mathbf{p}_{l+1},...,\mathbf{p}_n)\nonumber\\
&+ ... \,,
\label{decomposition}
\end{align}
where the final dots stay for all the terms with more than two annihilations\footnote{Clearly, the matrix elements $F_{m,n}^{\Phi}$ with $m$ and/or $n$ equal $0$ coincide with the connected ones.}.

As a consequence of (\ref{decomposition}), the expectation value (\ref{G1}) expands as
\begin{align}
G_\Phi(\mathbf{x},t) &=|f_0|^2 F^{\Phi}_{0,0}+\nonumber\\
&+\int \frac{d\mathbf{p}_1}{(2\pi)^d E_{\mathbf{p}_1}} f_1(\mathbf{p}_1) f_0^* e^{-i(\mathbf{p}_1\cdot \mathbf{x}+E_{\mathbf{p}_1}t)}F^{\Phi}_{0,1}(|\mathbf{p}_1)+c.c.\nonumber\\
&+\int \frac{d\mathbf{p}_1 d\mathbf{q}_1}{(2\pi)^{2d} E_{\mathbf{p}_1}E_{\mathbf{q}_1}} f_1(\mathbf{p}_1) f_1^*(\mathbf{q}_1) e^{i[(\mathbf{q}_1-\mathbf{p}_1)\cdot \mathbf{x}+(E_{\mathbf{q}_1}-E_{\mathbf{p}_1})t]} F^{\Phi,c}_{1,1}(\mathbf{q}_1|\mathbf{p}_1)\nonumber\\
&+ \int \frac{d\mathbf{p}_1}{(2\pi)^d E_{\mathbf{p}_1}} |f_1(\mathbf{p}_1)|^2 F^{\Phi}_{0,0}\nonumber\\
&+ \int \frac{d\mathbf{p}_1 d\mathbf{p}_2}{(2\pi)^{2d} E_{\mathbf{p}_1} E_{\mathbf{p}_2}} f_2(\mathbf{p}_1,\mathbf{p}_2) f_0^* e^{-i(\mathbf{P}\cdot \mathbf{x}+Et)} F^{\Phi}_{0,2}(|\mathbf{p}_1,\mathbf{p}_2) + c.c.\nonumber\\
&+\int \frac{d\mathbf{p}_1 d\mathbf{p}_2 d\mathbf{q}_1}{(2\pi)^{3d} E_{\mathbf{p}_1} E_{\mathbf{p}_2} E_{\mathbf{q}_1}} f_2(\mathbf{p}_1,\mathbf{p}_2) f_1^*(\mathbf{q}_1) e^{i[(\mathbf{q}_1-\mathbf{P})\cdot \mathbf{x} + (E_{\mathbf{q}_1}-E)t]} F^{\Phi,c}_{1,2}(\mathbf{q}_1|\mathbf{p}_1,\mathbf{p}_2) + c.c.\nonumber\\
&+2\int \frac{d\mathbf{p}_1 d\mathbf{p}_2}{(2\pi)^{2d} E_{\mathbf{p}_1} E_{\mathbf{p}_2}} f_2(\mathbf{p}_1,\mathbf{p}_2) f_1^*(\mathbf{p}_2) e^{-i(\mathbf{p}_1\cdot \mathbf{x} +E_{\mathbf{p}_1}t)} F^{\Phi}_{0,1}(|\mathbf{p}_1) + c.c.\nonumber\\
&+ \int \frac{d\mathbf{p}_1 d\mathbf{p}_2 d\mathbf{q}_1 d\mathbf{q}_2}{(2\pi)^{4d} E_{\mathbf{p}_1} E_{\mathbf{p}_2} E_{\mathbf{q}_1} E_{\mathbf{q}_2}} f_2(\mathbf{p}_1,\mathbf{p}_2) f_2^*(\mathbf{q}_1,\mathbf{q}_2) e^{i[(\mathbf{Q}-\mathbf{P})\cdot \mathbf{x}+(\tilde{E}-E)t]} F^{\Phi,c}_{2,2}(\mathbf{q}_1,\mathbf{q}_2|\mathbf{p}_1,\mathbf{p}_2)\nonumber\\
&+4 \int \frac{d\mathbf{p}_1 d\mathbf{p}_2 d\mathbf{q}_1}{(2\pi)^{3d} E_{\mathbf{p}_1} E_{\mathbf{p}_2} E_{\mathbf{q}_1}} f_2(\mathbf{p}_1,\mathbf{p}_2) f_2^*(\mathbf{q}_1,\mathbf{p}_2) e^{i[(\mathbf{q}_1-\mathbf{P})\cdot \mathbf{x}+(E_{\mathbf{q}_1}-E)t]} F^{\Phi,c}_{1,1}(\mathbf{q}_1|\mathbf{p}_1)\nonumber\\
&+2 \int \frac{d\mathbf{p}_1 d\mathbf{p}_2}{(2\pi)^{2d} E_{\mathbf{p}_1} E_{\mathbf{p}_2}} |f_2(\mathbf{p}_1,\mathbf{p}_2)|^2 F^{\Phi}_{0,0}+\cdots\,,
\label{G2}
\end{align}
where the complex conjugated ($c.c.$) terms come from the relation
\begin{equation}
F^{\Phi,c}_{n,m}(\mathbf{p}_1,...,\mathbf{p}_n|\mathbf{q}_1,...,\mathbf{q}_m)=[F^{\Phi,c}_{m,n}(\mathbf{q}_1,...,\mathbf{q}_m|\mathbf{p}_1,...,\mathbf{p}_n)]^*\,
\label{F_conj}
\end{equation}
satisfied by the Hermitian operators we consider. It follows that the expansion (\ref{G1}) can be re-expressed in terms of the connected matrix elements as 
\begin{align}
G_\Phi(\mathbf{x},t) &=\sum_{n,m=0}^{\infty}\int\prod_{i=1}^n \frac{d\mathbf{p}_i}{(2\pi)^d E_{\mathbf{p}_i}}\, \prod_{j=1}^m \frac{d\mathbf{q}_j}{(2\pi)^d E_{\mathbf{q}_j}}\,g_{m,n}(\mathbf{q}_1,...,\mathbf{q}_m,\mathbf{p}_1,...,\mathbf{p}_n)\nonumber\\
&\times F_{m,n}^{\Phi,c}(\mathbf{q}_1,...,\mathbf{q}_m|\mathbf{p}_1,...,\mathbf{p}_n)\,e^{i[(\mathbf{Q}-\mathbf{P})\cdot \mathbf{x}+(\tilde{E}-E)t]}\,,
\label{G3}
\end{align}
where the coefficient functions $g_{m,n}(\mathbf{q}_1,...,\mathbf{q}_m,\mathbf{p}_1,...,\mathbf{p}_n)$ 
%satisfy 
%\EQ
%g_{n,m}(\mathbf{p}_1,...,\mathbf{p}_n,\mathbf{q}_1,...,\mathbf{q}_m)=[g_{m,n}(\mathbf{q}_1,...,\mathbf{q}_m,\mathbf{p}_1,...,\mathbf{p}_n)]^*\,,
%\label{g_conj}
%\EN
expand in terms of the coefficients of (\ref{G1}) as
\begin{align}
    &g_{m,n}(\mathbf{q}_1,...,\mathbf{q}_m,\mathbf{p}_1,...,\mathbf{p}_n)= \nonumber\\
    &\sum_{k=0}^{\infty} \frac{(m+k)!\, (n+k)!}{m!\, n!\, k!} \int \prod_{i=1}^k \frac{d\mathbf{a}_i}{(2\pi)^d E_{\mathbf{a}_i}}\, f^*_{m+k}(\mathbf{q}_1,...,\mathbf{q}_m,\mathbf{a}_1,...,\mathbf{a}_k)\, f_{n+k}(\mathbf{p}_1,...,\mathbf{p}_n,\mathbf{a}_1,...,\mathbf{a}_k)\,.
\end{align}

Notice that the contribution to (\ref{G3}) with $m=n=0$ has, in particular, $E=\tilde{E}=0$, and is then time-independent. If we subtract it from (\ref{G3}) defining
\begin{align}
G^s_\Phi(\mathbf{x},t)&=\sum\limits_{\substack{m,n=0\\(m,n)\neq(0,0)}}^{\infty}\int\prod_{i=1}^n \frac{d\mathbf{p}_i}{(2\pi)^d E_{\mathbf{p}_i}}\, \prod_{j=1}^m \frac{d\mathbf{q}_j}{(2\pi)^d E_{\mathbf{q}_j}}\,g_{m,n}(\mathbf{q}_1,...,\mathbf{q}_m,\mathbf{p}_1,...,\mathbf{p}_n)\nonumber\\
&\times F_{m,n}^{\Phi,c}(\mathbf{q}_1,...,\mathbf{q}_m|\mathbf{p}_1,...,\mathbf{p}_n)\,e^{i[(\mathbf{Q}-\mathbf{P})\cdot \mathbf{x}+(\tilde{E}-E)t]}\,,
\label{Gs}
\end{align}
the one-point function (\ref{continuity}) can be rewritten as
\EQ
\langle\Phi(\mathbf{x},t)\rangle=G^s_\Phi(\mathbf{x},t)-G^s_\Phi(\mathbf{x},0)+\langle\Phi(\mathbf{x},0)\rangle\,,
\hspace{1cm}t\geq 0\,.
\label{continuity_s}
\EN

\section{Translation invariant case}
So far we allowed for $t>0$ the presence of spatial inhomogeneities inherited from the time evolution at negative times. We now consider the case in which such inhomogeneities are absent and the state of the system is translation invariant. This is a particular case in which the coefficient functions in (\ref{state}) take the form\footnote{For $n=0$ there are no momenta and (\ref{factorization}) reduces to $f_0=\hat{f}_0$.} 
\EQ
f_n(\mathbf{p}_1,...,\mathbf{p}_n)=\delta(\mathbf{P})\,\hat{f}_n(\mathbf{p}_1,...,\mathbf{p}_n)\,,
\label{factorization}
\EN
so that we have
\begin{equation}
|\psi\rangle = \sum_{n=0}^{\infty}\int_{-\infty}^{\infty} \prod_{i=1}^n \frac{d\mathbf{p}_i}{(2\pi)^d E_{\mathbf{p}_i}}\, \delta(\mathbf{P})\, \hat{f}_n(\mathbf{p}_1,...,\mathbf{p}_n)\, |\mathbf{p}_1,...,\mathbf{p}_n\rangle\,,
\label{state_t_i}
\end{equation}
and the expression (\ref{G1}) for the expectation value becomes
\begin{align}
G_\Phi(t) &=\sum_{n,m=0}^{\infty}\int\prod_{i=1}^n \frac{d\mathbf{p}_i}{(2\pi)^d E_{\mathbf{p}_i}}\, \prod_{j=1}^m \frac{d\mathbf{q}_j}{(2\pi)^d E_{\mathbf{q}_j}}\,\delta(\mathbf{P}) \delta(\mathbf{Q})\nonumber\\
&\times \hat{f}_n(\mathbf{p}_1,...,\mathbf{p}_n)\, \hat{f}^*_m(\mathbf{q}_1,...,\mathbf{q}_m)\, F_{m,n}^{\Phi}(\mathbf{q}_1,...,\mathbf{q}_m|\mathbf{p}_1,...,\mathbf{p}_n)\, e^{i(\tilde{E}-E)t}\,.
\label{Gti}
\end{align}
Use of (\ref{decomposition}) now leads to an expansion in terms of connected matrix elements which for the subtracted expectation value (\ref{Gs}) reads
\begin{align}
G^s_\Phi(t) &=\int \frac{d\mathbf{p}_1}{(2\pi)^d E_{\mathbf{p}_1}} \delta(\mathbf{p}_1) \hat{f}_1(\mathbf{p}_1) \hat{f}_0^* e^{-iE_{\mathbf{p}_1}t}F^{\Phi}_{0,1}(|\mathbf{p}_1)+c.c.\nonumber\\
&+\int \frac{d\mathbf{p}_1 d\mathbf{q}_1}{(2\pi)^{2d} E_{\mathbf{p}_1}E_{\mathbf{q}_1}} \delta(\mathbf{p}_1) \delta(\mathbf{q}_1) \hat{f}_1(\mathbf{p}_1) \hat{f}_1^*(\mathbf{q}_1) e^{i(E_{\mathbf{q}_1}-E_{\mathbf{p}_1})t} F^{\Phi,c}_{1,1}(\mathbf{q}_1|\mathbf{p}_1)\nonumber\\
&+\int \frac{d\mathbf{p}_1 d\mathbf{p}_2}{(2\pi)^{2d} E_{\mathbf{p}_1} E_{\mathbf{p}_2}} \delta(\mathbf{p}_1+\mathbf{p}_2) \hat{f}_2(\mathbf{p}_1,\mathbf{p}_2) \hat{f}_0^* e^{-iEt} F^{\Phi}_{0,2}(|\mathbf{p}_1,\mathbf{p}_2) + c.c.\nonumber\\
&+\int \frac{d\mathbf{p}_1 d\mathbf{p}_2 d\mathbf{q}_1}{(2\pi)^{3d} E_{\mathbf{p}_1} E_{\mathbf{p}_2} E_{\mathbf{q}_1}} \delta(\mathbf{p}_1+\mathbf{p}_2)\delta(\mathbf{q}_1) \hat{f}_2(\mathbf{p}_1,\mathbf{p}_2) \hat{f}_1^*(\mathbf{q}_1) e^{i(E_{\mathbf{q}_1}-E)t} F^{\Phi,c}_{1,2}(\mathbf{q}_1|\mathbf{p}_1,\mathbf{p}_2) + c.c.\nonumber\\
&+2\int \frac{d\mathbf{p}_1}{(2\pi)^{2d} E_{\mathbf{p}_1} M} \delta(\mathbf{p}_1) \hat{f}_2(\mathbf{p}_1,0) \hat{f}_1^*(0) e^{-iE_{\mathbf{p}_1}t} F^{\Phi}_{0,1}(|\mathbf{p}_1) + c.c.\nonumber\\
&+\int \frac{d\mathbf{p}_1 d\mathbf{p}_2 d\mathbf{q}_1 d\mathbf{q}_2}{(2\pi)^{4d} E_{\mathbf{p}_1} E_{\mathbf{p}_2} E_{\mathbf{q}_1} E_{\mathbf{q}_2}} \delta(\mathbf{p}_1+\mathbf{p}_2)\delta(\mathbf{q}_1+\mathbf{q}_2) \hat{f}_2(\mathbf{p}_1,\mathbf{p}_2) \hat{f}_2^*(\mathbf{q}_1,\mathbf{q}_2) e^{i(\tilde{E}-E)t}\nonumber\\
&\times F^{\Phi,c}_{2,2}(\mathbf{q}_1,\mathbf{q}_2|\mathbf{p}_1,\mathbf{p}_2)\nonumber\\
&+4\int \frac{d\mathbf{p}_1 d\mathbf{q}_1}{(2\pi)^{3d} E_{\mathbf{p}_1} E_{\mathbf{q}_1}^2} \delta(\mathbf{q}_1-\mathbf{p}_1) \hat{f}_2(\mathbf{p}_1,-\mathbf{q}_1) \hat{f}_2^*(\mathbf{q}_1,-\mathbf{q}_1) e^{i(E_{\mathbf{q}_1}-E_{\mathbf{p}_1})t} F^{\Phi,c}_{1,1}(\mathbf{q}_1|\mathbf{p}_1)\nonumber\\
&+\cdots\,.
\end{align}
Notice that the terms coming from the connected part of the original matrix elements (\ref{matrixelement}) contain delta functions that do not mix the momenta $\{\mathbf{p}_i\}$ and $\{\mathbf{q}_j\}$, while the terms coming from the disconnected parts involve delta functions of differences of these momenta. Hence, the expansion in terms of the connected matrix elements can be written in the form
\begin{align}
G^s_\Phi(t) &=\sum\limits_{\substack{m,n=0\\(m,n)\neq(0,0)}}^{\infty}\int\prod_{i=1}^n \frac{d\mathbf{p}_i}{(2\pi)^d E_{\mathbf{p}_i}}\, \prod_{j=1}^m \frac{d\mathbf{q}_j}{(2\pi)^d E_{\mathbf{q}_j}}\, \big[ \delta(\mathbf{P}) \delta(\mathbf{Q})\, h_{m,n}(\mathbf{q}_1,...,\mathbf{q}_m,\mathbf{p}_1,...,\mathbf{p}_n)\nonumber\\
&+ \delta(\mathbf{Q}-\mathbf{P})\, \tilde{h}_{m,n}(\mathbf{q}_1,...,\mathbf{q}_m,\mathbf{p}_1,...,\mathbf{p}_n) \big]\, F_{m,n}^{\Phi,c}(\mathbf{q}_1,...,\mathbf{q}_m|\mathbf{p}_1,...,\mathbf{p}_n)\,e^{i(\tilde{E}-E)t},
\label{Gs2}
\end{align}
with coefficient functions $h_{m,n}(\mathbf{q}_1,...,\mathbf{q}_m,\mathbf{p}_1,...,\mathbf{p}_n)$ and $\tilde{h}_{m,n}(\mathbf{q}_1,...,\mathbf{q}_m,\mathbf{p}_1,...,\mathbf{p}_n)$ 
%satisfying 
%\EQ
%h_{n,m}(\mathbf{p}_1,...,\mathbf{p}_n,\mathbf{q}_1,...,\mathbf{q}_m)=[h_{m,n}(\mathbf{q}_1,...,\mathbf{q}_m,\mathbf{p}_1,...,\mathbf{p}_n)]^*\,,
%\label{h_conj}
%\EN
%\EQ
%\tilde{h}_{n,m}(\mathbf{p}_1,...,\mathbf{p}_n,\mathbf{q}_1,...,\mathbf{q}_m)=[\tilde{h}_{m,n}(\mathbf{q}_1,...,\mathbf{q}_m,\mathbf{p}_1,...,\mathbf{p}_n)]^*\,,
%\label{htilde_conj}
%\EN
related to the coefficients $\hat{f}_n$ of (\ref{Gti}) as 
\EQ
h_{m,n}(\mathbf{q}_1,...,\mathbf{q}_m,\mathbf{p}_1,...,\mathbf{p}_n)=
\hat{f}_n(\mathbf{p}_1,...,\mathbf{p}_n)\hat{f}_m^*(\mathbf{q}_1,...,\mathbf{q}_m)\,,
\label{hmn}
\EN
and\footnote{For $m=0$ and $k=1$ there no momenta ${\bf q}_i$ and ${\bf a}_i$. In this case $E_{\mathbf{A}+\mathbf{Q}}=E_0=M$.}
\begin{align}
    &\tilde{h}_{m,n}(\mathbf{q}_1,...,\mathbf{q}_m,\mathbf{p}_1,...,\mathbf{p}_n)=\sum_{k=1}^{\infty} \frac{(m+k)!\, (n+k)!}{m!\,n!\, k!\, (2\pi)^d} \int \prod_{i=1}^{k-1} \frac{d\mathbf{a}_i}{(2\pi)^d E_{\mathbf{a}_i}}\, \frac{1}{E_{\mathbf{A}+\mathbf{Q}}}\nonumber\\
    &\times \hat{f}_{n+k}(\mathbf{p}_1,...,\mathbf{p}_n,\mathbf{a}_1,...,\mathbf{a}_{k-1},-\mathbf{A}-\mathbf{Q})\, \hat{f}^*_{m+k}(\mathbf{q}_1,...,\mathbf{q}_m,\mathbf{a}_1,...,\mathbf{a}_{k-1},-\mathbf{A}-\mathbf{Q})\,,
\label{htilde}    
\end{align}
where
\begin{equation}
    \mathbf{A}=\sum_{i=1}^{k-1} \mathbf{a}_i\,.
\end{equation}

For $t\to\infty$ the integrand of (\ref{Gs2}) rapidly oscillates because of the exponential factor $e^{i(\tilde{E}-E)t}$ and suppresses the integrals unless the phase is stationary, namely unless the momenta are small. The coefficient functions and the matrix elements in (\ref{Gs2}) generically go to constants in this limit\footnote{The case $d=1$ involves some additional consideration that we postpone to section \ref{pt}.}. The behavior for $t$ large enough of the contribution to (\ref{Gs2}) with $(m,n)$ quasiparticles is then obtained using the nonrelativistic expression of the energies and rescaling the momentum components by $\sqrt{t}$; this gives
\EQ
\bigg[B_{m,n}^{\Phi}\,t^{-(m+n-2+\delta_{n,0}+\delta_{m,0})d/2}+\tilde{B}_{m,n}^{\Phi}\,t^{-(m+n-1-\delta_{n,1}\delta_{m,1})d/2} \bigg]\, e^{i(m-n)Mt},
\label{large_time}
\EN 
where $B_{m,n}^{\Phi}$ and $\tilde{B}_{m,n}^{\Phi}$ are constants and we took into account that the term $(m,n)=(0,0)$ is not present in (\ref{Gs2}). We see that the leading time dependence comes from $(m,n)$ equal $(1,0)$ and $(0,1)$ and corresponds to undamped oscillations. Notice that in absence of the delta function in (\ref{state_t_i}) (i.e. in the generic inhomogeneous case of previous section) the oscillations coming from the $(1,0)$ and $(0,1)$ contributions would be damped as $t^{-d/2}$. 

Since the only relativistic invariant that can be formed from the energy and momentum of a single particle is a constant, $F^\Phi_{m,n}$ with $m+n=1$ is a constant. Besides (\ref{F_conj}), the matrix elements $F^\Phi_{m,n}$ with $m$ and $n$ interchanged are related by crossing symmetry \cite{ELOP}, which amounts to analytic continuation in the momenta; this leads to the real constant 
\EQ
F^\Phi_{0,1}=F^\Phi_{1,0}\equiv F^\Phi_1\,.
\label{F1}
\EN
Putting all together, the large time limit of the one-point function (\ref{continuity_s}) is given by (\ref{asympt}) with
\EQ
h_1= h_{0,1}(0)+\tilde{h}_{0,1}(0)\,,
\label{h1}
\EN
and
\begin{align}
A_\Phi&=\langle\Phi(0)\rangle-\sum\limits_{\substack{m,n=0\\(m,n)\neq(0,0),(1,1)}}^{\infty}\int\prod_{i=1}^n \frac{d\mathbf{p}_i}{(2\pi)^d E_{\mathbf{p}_i}}\, \prod_{j=1}^m \frac{d\mathbf{q}_j}{(2\pi)^d E_{\mathbf{q}_j}}\, \big[ \delta(\mathbf{P}) \delta(\mathbf{Q})\nonumber\\
&\times h_{m,n}(\mathbf{q}_1,...,\mathbf{q}_m,\mathbf{p}_1,...,\mathbf{p}_n)+\delta(\mathbf{Q}-\mathbf{P})\, \tilde{h}_{m,n}(\mathbf{q}_1,...,\mathbf{q}_m,\mathbf{p}_1,...,\mathbf{p}_n) \big]\nonumber\\
&\times F_{m,n}^{\Phi,c}(\mathbf{q}_1,...,\mathbf{q}_m|\mathbf{p}_1,...,\mathbf{p}_n)\,.
\label{A_Phi}
\end{align}

In the physical dynamical problems we consider, the $\hat{f}_n$'s in (\ref{state_t_i}) are nonzero unless an internal symmetry\footnote{The symmetry can also be topological, see the example of \cite{q_int}.} forces some of them to vanish. In the current case of a single particle species, a $\mathbb{Z}_2$ symmetry may lead to the vanishing of the $\hat{f}_n$'s with $n$ even or of those with $n$ odd. Since $h_1$ is a sum of terms containing $\hat{f}_n^*\hat{f}_{n+1}$, we have $h_1\neq 0$ unless $\hat{f}_0\hat{f}_{1}=0$. Hence, the condition for the presence of undamped oscillations in (\ref{asympt}) is $\hat{f}_0\hat{f}_{1}F_1^\Phi\neq 0$. 

Notice that the asymptotic offset (\ref{A_Phi}) differs from the constant $\langle\Phi(0)\rangle-G^s_\Phi(0)$ entering (\ref{continuity_s}) for the subtraction of the term $(m,n)=(1,1)$ in the sum; the reason is that in (\ref{continuity_s}) this term is canceled by the $(1,1)$ contribution to (\ref{Gs2}), which has $\tilde{E}=E$ and is time-independent. Equation (\ref{large_time}) shows that the first subleading contributions to (\ref{asympt}) come from $(m,n)$ equal $(0,2)$, $(2,0)$, $(1,2)$ and $(2,1)$ and correspond to damped oscillations. 

The analysis we performed above is easily generalized along the same lines to the case of several quasiparticle species $a=1,2,...,k$ with masses $M_a$. In particular, the oscillations that remain undamped at large times take the form
\EQ
\sum_{a=1}^k \frac{F_{1a}^{\Phi}}{(2\pi)^d M_a}(h_{1a}\,e^{-iM_at}+h^*_{1a}\,e^{iM_at}) 
+\sum\limits_{\substack{a,b\\M_a\neq M_b}} \frac{F_{1b,1a}^{\Phi}(0|0)}{(2\pi)^{2d}} \frac{h_{1b,1a}(0,0)}{M_a M_b} e^{i(M_b-M_a)t},
\label{ab}
\EN
where the first sum generalizes the term present in (\ref{asympt}), with $F_{1a}^{\Phi}$ and $h_{1a}$ corresponding to (\ref{F1}) and (\ref{h1}) with the specification of the species of the particle. The second sum, on the other hand, is a contribution arising from the fact that the term $(m,n)=(1,1)$ is no longer time-independent when the two particles have different masses; $h_{1b,1a}$ corresponds to $h_{1,1}$ of (\ref{hmn}) with the specification of the quasiparticle species\footnote{For $a\neq b$, $F_{1b,1a}^{\Phi}=F_{1b,1a}^{\Phi,c}$ follows from the fact that particles of different species cannot annihilate each other. The contribution multiplying $\tilde{h}_{1b,1a}$ is damped at large times.}. We also see that the condition for the presence of this second type of undamped oscillations is $\hat{f}_{1b}\hat{f}_{1a}F_{1b,1a}^\Phi\neq 0$. Once again this condition involves one-quasiparticle states and is satisfied unless an internal symmetry causes the vanishing of one of the three factors. This clarifies the role of symmetries for undamped oscillations, a role that had been debated in the literature (see \cite{MBJ} and references therein). In the perturbative theory of quantum quenches, the undamped oscillations with frequencies $M_b-M_a$ arise at second order in the quench size \cite{oscill}, as will also be seen in the next section.

%It is worth mentioning that in numerical simulations dynamically generated nonequilibrium is sometimes traded for time evolution from some engineered initial condition at $t=0$, and in principle some coefficients ${f}_n$ may be set to zero by hand, indepenently of symmetry considerations. In this case our formulae continue to hold with the set of ${f}_n$'s selected in this way.

\section{Comparison with perturbative results}
\label{pt}
The Hamiltonian (\ref{Hamiltonian}) includes as a particular case that in which the negative and positive time Hamiltonians differ for the change of an interaction parameter, namely the homogeneous quench
\EQ
\left\{
\begin{array}{l}
H_0\,,\hspace{4.4cm}t<0\,,\\
\\
H=H_0+\lambda\int_{-\infty}^\infty d{\bf x}\,\Psi({\bf x})\,,\hspace{.7cm}t\geq 0\,.
\end{array}
\right.
\label{quench}
\EN
We then refer to $\Psi({\bf x})$ as the quench operator and to $\lambda$ as the quench size. A general perturbative analysis can be performed in the quench size \cite{quench}, in any dimension $d$ \cite{oscillD} and for arbitrarily strong interaction among the quasiparticles. When the system is in the ground state of $H_0$ for negative times, the post-quench state is given by (\ref{state_t_i}) with\footnote{The result (\ref{pert_n}) shows the peculiarity of the case of noninteracting quasiparticles, for which $H_0$ and the quench operator are quadratic in the excitation modes and $F^{\Psi}_{n,0}$ vanishes for $n\neq 2$. As a consequence the post-quench state is made of pairs of quasiparticles with opposite momenta, a structure that does not occur for interacting quasiparticles. \label{pair}} \cite{quench,oscillD}
\begin{align}
&\hat{f}_0=1+O(\lambda^2)\,,
\label{pert_1}\\
&\hat{f}_{n\ge 1}(\mathbf{p}_1,...,\mathbf{p}_n)=\lambda\, \frac{(2\pi)^d}{n! E}\, F^{\Psi}_{n,0}(\mathbf{p}_1,...,\mathbf{p}_n|)+O(\lambda^2)\,.
\label{pert_n}
\end{align}
It then follows from (\ref{h1}), (\ref{hmn}), (\ref{htilde}) and (\ref{F1}) that
\EQ
h_1=h_{0,1}(0)+O(\lambda^2)=\hat{f}_1(0)+O(\lambda^2)=\frac{\lambda}{M} (2\pi)^d F_1^{\Psi} + O(\lambda^2)\,,
\EN
and from (\ref{asympt}) that undamped oscillations
\EQ
\lambda \,\frac{2}{M^2} F_1^{\Psi} F_1^{\Phi} \cos Mt+O(\lambda^2)
\label{oscill_pert}
\EN 
show up already at leading order in the quench size, as originally shown in \cite{quench}. Notice also that (\ref{hmn}) leads to
\EQ
h_{1b,1a}(0,0)=\lambda^2\,\frac{(2\pi)^{2d}}{M_aM_b}\,F_{1a}^\Psi F_{1b}^\Psi+O(\lambda^3)\,,
\label{ab_pert}
\EN 
so that in perturbation theory the undamped oscillations with frequences $M_a-M_b$ in (\ref{ab}) arise at second order in the quench size, as observed in \cite{oscill}. 

While the expressions (\ref{oscill_pert}) and (\ref{ab_pert}) coincide with those of the perturbative calculations of \cite{quench,oscill}, a subtlety has to be pointed out: those perturbative calculations were performed in the basis of the quasiparticle states of the pre-quench theory (i.e. the unperturbed theory $\lambda=0$), while the basis we use in this paper is that of the $t>0$ theory. The point, however, is that the difference between the two bases can be ignored when working at leading order in perturbation theory\footnote{This is true in the generic case for which the quasiparticle content does not change when passing from $\lambda=0$ to $\lambda$ small. See \cite{DV} for the discussion and examples of the special case in which this condition is not fulfilled.}.

We see from (\ref{pert_1}), (\ref{pert_n}), (\ref{hmn}) and (\ref{htilde}) that $h_{m,n}$ are of order $\lambda^2$ unless $m$ or $n$ vanish, and that $\tilde{h}_{m,n}$ are in any case of order $\lambda^2$. It follows that the asymptotic offset (\ref{A_Phi}) takes the form
\begin{align}
A_\Phi&=\langle\Phi(0)\rangle-\sum_{n=1}^{\infty}\int\prod_{i=1}^n \frac{d\mathbf{p}_i}{(2\pi)^d E_{\mathbf{p}_i}}\, \delta(\mathbf{P}) \big[h_{0,n}(\mathbf{p}_1,...,\mathbf{p}_n)\,F_{0,n}^{\Phi}(|\mathbf{p}_1,...,\mathbf{p}_n)+c.c.\big]\nonumber\\
&+O(\lambda^2)\,,
\label{A_Phi_pert}
\end{align}
where $\langle\Phi(0)\rangle$ is now the expectation value on the ground state of the pre-quench theory, and
\EQ
h_{0,n}(\mathbf{p}_1,...,\mathbf{p}_n)=\hat{f}_0^*\hat{f}_n(\mathbf{p}_1,...,\mathbf{p}_n)=
\lambda\, \frac{(2\pi)^d}{n! E}\, [F^{\Psi}_{0,n}(|\mathbf{p}_1,...,\mathbf{p}_n)]^*+O(\lambda^2)\,.
\label{hn_pert}
\EN
It was shown in \cite{DV} that (\ref{A_Phi_pert}), (\ref{hn_pert}) lead to
\EQ
A_\Phi=\langle \Phi\rangle_{\lambda}+O(\lambda^2)\,,
\EN 
where $\langle \Phi\rangle_{\lambda}$ is the expectation value on the ground state of the post-quench theory. The nonperturbative expression (\ref{A_Phi}) suggests that in general there is no simple way of expressing the offset. 

In our analysis of the large time behavior of one-point functions in the previous section we used the fact that the matrix elements (\ref{connectedmatrixelement}) generically go to some finite constant when the momenta tend to zero. In $d=1$, however, the quasiparticles often possess fermionic statistics\footnote{The basic example is provided by the Ising chain, see \cite{review} for a review.}, and for the matrix elements (\ref{connectedmatrixelement}) this leads to a zero when ${\bf q}_i={\bf q}_j$ or ${\bf p}_i={\bf p}_j$, and to a pole when ${\bf q}_i={\bf p}_j$. The poles are accompanied by an $i\epsilon$ prescription \cite{Smirnov} which anyway displaces them from the integration path over momenta\footnote{See \cite{q_int} for a basic and physically interesting illustration of this feature in the nonequilibrium context.}. The effect of the zeros in the matrix elements (\ref{connectedmatrixelement}) and in the coefficient functions which multiply them in the expressions such as (\ref{Gs2}) is to enhance the time decay in (\ref{large_time}), without affecting the undamped oscillations in (\ref{asympt}) which are generally derived in any dimension. The perturbative results recalled above are of course consistent with this fact, and indicate that fermionic statitistics in $d=1$ enhances the decay of the remainder in (\ref{asympt}) ($t^{-3/2}$ instead of $t^{-1/2}$) \cite{quench,oscillD}. If the quasiparticles have fermionic statistics the suitable sign factors have to be introduced in (\ref{decomposition}) and will affect the combinatorial prefactor in (\ref{htilde}).

\section{Conclusion}
In this paper we studied the nonequilibrium dynamics of quantum statistical systems in $d$ spatial dimensions which for positive times evolve with a Hamiltonian $H$ which is time-independent and translation invariant in space. The nonequilibrium state was expanded on the basis of energy eigenstates (asymptotic quasiparticle states) of $H$, with coefficient functions $f_n$ which were left generic in order to account for arbitrary evolution for $t<0$ under some Hamiltonian $H_0({\bf x},t)$. We then showed how the evolution for positive times of the one-point functions of local operators (e.g. the order parameter) depends on the $f_n$'s. 

While the theory shows that the large time dynamics is determined by low-energy modes, our framework ensures that the results hold also in the vicinity of quantum critical points. It also allows to appreciate the role played by the connectedness structure of matrix elements, a circumstance noted since \cite{lightcone}, where this structure was shown to account for the light cone spreading of correlations in two-point functions. 

In the generic case (\ref{Hamiltonian}), in which translation invariance is absent, the theory leads to oscillations of the one-point functions that normally decay as $t\to\infty$. This is the case, in particular, when the system is confined in a finite region of space before this spatial constraint is removed for $t>0$ (release from a trap\footnote{See the early experimental realization of \cite{Kinoshita}, where oscillations were observed.}). In such a situation, the energy density carried by the quasiparticle excitations goes to zero as $t\to\infty$ (local dissipation) and is insufficient to sustain the oscillations at large enough times. A first illustration of this phenomenon was given perturbatively in \cite{oscill,oscillD} in the framework of inhomogeneous quantum quenches.

On the other hand, when the analysis is specialized to the case in which no spatial inhomogeneity is inherited from negative times, the theory shows that one-point functions exhibit undamped oscillations when no internal symmetry prevents a one-quasiparticle contribution to the nonequilibrium state or the coupling of the operator to this contribution. This result confirms the one obtained perturbatively since \cite{quench,oscill} for the case of the instantaneous change of an interaction parameter. 
%Since the theory shows the role of translation invariance in keeping the oscillations undamped, observing no damping in the cases predicted by the theory can be used to test up to which timescale numerical simulations remain reliable (i.e. insensitive to finite size or other undesired effects). 

We also obtained the expression (\ref{A_Phi}) of the asymptotic offset of one-point functions in terms of the matrix elements of the operator and of the coefficients specifying the nonequilibrium state. We showed how the structure of this result simplifies in the particular case of a small quench from the ground state and allows, up to higher order corrections in the quench size, the resummation originally shown in \cite{DV}. While no similar resummations seem likely for the full result (\ref{A_Phi}), it is known that expansions over quasiparticles modes often converge rapidly providing very good approximations from the first few terms\footnote{This has been checked in detail for integrable models, see \cite{review}.}. It will be interesting to investigate to which extent this happens in the present case.

\appendix
\section{Appendix}
Since energy is conserved, nothing can force collective oscillation modes of an isolated homogeneous statistical system to always decay. In this respect, the undamped oscillating term (\ref{oscill_pert}) obtained in \cite{quench} for one-point functions after a quench of size $\lambda$ provided the first analytical result. The fact that $F^\Psi_1$ vanishes for noninteracting quasiparticles naturally accounted for the circumstance that the undamped oscillations were not found in \cite{BMcD} nor in the many studies devoted in more recent years to the transverse field Ising chain (free fermions). The result (\ref{oscill_pert}) also explained that interaction alone is not sufficient to produce such oscillations; it is also necessary that no internal symmetry causes the vanishing of the one-quasiparticle matrix elements $F_1^\Psi$ and $F_1^\Phi$. It was then pointed out in \cite{quench} that the simplest system where to look for those oscillations is the Ising chain with longitudinal field. Indeed, the longitudinal field introduces interaction among the quasiparticles and leaves no internal symmetry. The author of \cite{quench} did not know that "strong numerical evidence" of undamped oscillations had already been observed precisely in this model in \cite{BCH}.

The analysis of \cite{quench} was perturbative in the quench size $\lambda$, and it was pointed out in the same paper that the results of finite order perturbation theory can be quantitatively accurate up to the timescale $t_\lambda\sim 1/\lambda^{1/(d+1-X_\Psi)}$, where $X_\Psi$ is the scaling dimension of the quench operator. While $t_\lambda$ can be made arbitrarily large reducing the size of the quench, it is finite for fixed $\lambda$ and the question of the fate of the oscillations beyond this timescale has to be posed. It was then observed in \cite{oscill} that, for $F_1^{\Psi} F_1^{\Phi}\neq 0$, the first order result (\ref{oscill_pert}) actually implies undamped oscillations in the full result (first order plus higher orders) for the one-point functions of usual physical interest. To see this, let us focus on the remainder of (\ref{oscill_pert}), namely the resummation of all terms beyond first order. Mathematically, this will be a function that for $t\to\infty$ can either diverge, or approach a constant value, or itself exhibit undamped oscillations. On the other hand, since the contribution of order $\lambda$ is bounded, the first possibility cannot occur for ordinary physical observables such as a local magnetization: at a given point of space this can grow in time at most to the limit of maximal ordering, but cannot diverge. Having discarded the possibility of divergence, the other two possibilities lead to undamped oscillations for the complete result (first order plus remainder) as long as these are present at first order\footnote{Clearly, this observation does not imply that the oscillations of the complete result are those of the first order.} \cite{oscill}. 

What we just recalled clearly indicated that it should be possible to derive the presence of undamped oscillations of one-point functions in isolated homogeneous systems with interacting quasiparticles without reference to perturbation theory, and this is what we have done in this paper. The result (\ref{asympt}) is mathematically clear: for $t\to\infty$ the remainder vanishes at least as $t^{-d/2}$ and there will be undamped oscillations as long as $h_1F_1^\Phi\neq 0$. While in this respect this is the end of the story, it may be worth spending some words on a point that might be confusing. Before \cite{quench} there was no way to analytically determine the state produced by the quench (\ref{quench}) in the generic case of interacting quasiparticles and some studies based on trial states were proposed. In particular, with the idea that this could be relevant for models with interacting quasiparticles and integrable equilibrium dynamics in $d=1$, it was proposed (see e.g. \cite{GDLG}) to consider states {\it of the kind} arising in a different physical problem, namely that of perfectly reflecting boundary conditions in integrable models \cite{GZ}. These states are made of pairs of quasiparticles with opposite momenta and this "pair structure" exponentiates. Hence, we refer to them as exponential states; sometimes they are also called "integrable states" for the analogy we recalled. Their relevance for the quench problem (\ref{quench}) with interacting quasiparticles could not be shown and, in the words of \cite{BSE}, this issue was "simply set aside". On the other hand, the results of \cite{quench} recalled in section~\ref{pt} (remember footnote \ref{pair}) showed that such a peculiar feature as the pair structure -- not to mention its exponentiation -- does not arise in the quench (\ref{quench}) if the quasiparticles interact, and that in such a case the full complexity of the state (\ref{state_t_i}) needs to be faced\footnote{The values of the coefficient functions $f_n({\bf p}_1,\ldots,{\bf p}_n)$ allowed by the exponentiated pair structure form a zero measure subset of the total.}. 

Hence, we see that in the case of interacting quasiparticles these exponential states can only be interpreted as initial conditions in which at $t=0$ the state is fine tuned by hand on the exponentiated pair structure; they do not arise in our problem (\ref{Hamiltonian}) in which nonequilibrium is generated dynamically by the time evolution since $t=-\infty$ and no fine tuning is possible. This essential physical difference completely changes also the technical nature of the problem with respect to our case. Indeed, since an exponential state does not allow the notions of negative time and of pre-quench state\footnote{In particular, the notion of continuity of one-point functions at $t=0$ also disappears.}, one-point functions are normalized dividing by the norm of the exponential state itself. Divergences specifically due to the fine tuning on the pair structure appear both in the numerator and denominator, and a "linked cluster expansion" (in the amplitude of the quasiparticle pairs assumed small) as well as specific regularization steps are devised in order to produce a finite result (see \cite{BSE,CS,HKT}). This recipe yields some positive powers of time retaining memory of the assumed exponential structure of the state, and these powers are conjectured to resum into an exponential. In this way an exponential damping of one-quasiparticle oscillation modes was proposed in \cite{CS}. However, we see that this exponential damping of oscillations is a conjecture relying on a chain of assumptions, and that already the first of these assumptions (the pair structure) is ruled out in the problem of dynamically generated nonequilibrium with interacting quasiparticles considered in the present paper.

It may also be worth pointing out that this example of exponential states illustrates something more general. The derivation of the present paper shows that the possibility of undamped collective oscillation modes follows from general properties of unitary time evolution in isolated homogeneous quantum systems in any dimension, as required by the fact that nothing can force oscillation modes to always decay in presence of energy conservation. {\it No fine tuning} of the coefficient functions $f_n({\bf p}_1,\ldots,{\bf p}_n)$ in the state (\ref{state}) to some special form can change the conclusion valid for the generic case. In addition, no such fine tuning can be physically justified for the class (\ref{Hamiltonian}) -- or its subclass (\ref{quench}) -- of dynamical problems we considered. 

The irrelevance of fine tuning arguments for the problem at hand is by now well known and they are not mentioned in \cite{RSE}, a recent paper considering persistent oscillations. These authors, clearly unaware that the observation of \cite{oscill} recalled earlier in this appendix rules out this possibility, asked whether oscillations undamped at first order in the quench size can altogether disappear in the full result. It is then not surprising that their treatment by a truncated BBGKY hierarchy of a dimerized XXZ chain in a staggered magnetic field was inconclusive: in their words, "it is not impossible that higher-order corrections would cause the oscillations to remain at late times"\footnote{It is worth stressing that the perturbative parameter involved in the considerations of \cite{RSE} is not the quench size of \cite{quench} but the interaction strength. This is problematic, since the oscillations in their model are attributed to a bound state. The latter is nonperturbative in the interaction strength and a consistent perturbative study of the oscillations starting from the noninteracting theory is awkward in principle before than in practice. This problem is absent in the perturbation theory in the quench size, where the quasiparticle spectrum is already that of the interacting theory and can include bound states, topological excitations, etc. \cite{quench,DV,oscillD}.}. Of course, from the general point of view, the possibility of undamped oscillations is not questioned in \cite{RSE} and realizations are listed\footnote{See the published version.}. In the present paper we showed nonperturbatively how and under which conditions they emerge.

Since undamped oscillation modes are clearly possible in energy-conserving homogeneous systems and are theoretically understood, it is not surprising that they are observed in numerical simulations. As a matter of fact, "strong numerical evidence that such states do not relax even at very long times" was obtained already several years ago \cite{BCH} for the Ising chain with longitudinal field\footnote{From the point of view of the historical development of numerical studies, it is interesting to recall that the perturbative results of \cite{quench} (amplitude and frequency of the oscillations) were first numerically confirmed in \cite{RMCKT} for a quench of the longitudinal field in the Ising chain. A decrease of the amplitude in time was observed in the same paper for a double quench (i.e. both in the longitudinal and transverse field), for which no perturbative formula is available. Since only the first three oscillations were observed, this initial decrease did not imply a persistence of the damping at larger times. And indeed, no persistence of the damping was observed in \cite{KCTC}, where the effects of a double quench were followed over twenty oscillations.}. For its simplicity, this model minimizes the numerical efforts and leaves no margins of interpretation about the timescales. Outstanding evidence on the absence of damping was then given in \cite{Jacopo}, where the time evolution was followed over hundreds of oscillations and times exceeding by {\it three orders of magnitude} the two timescales ($t_\lambda$ and the inverse mass gap) allowed by the analytical formulation of the quench\footnote{The proposal of \cite{BBK} that a further timescale might exist beyond which the oscillations might decay appears of difficult verification. Concerning \cite{Jacopo}, the study includes the entanglement entropy. The latter corresponds to the one-point function of the twist field \cite{CC2} and, as far the oscillations are concerned, falls into the predictive domain of the general theory for one-point functions. The authors simulated also the longitudinal magnetization and equally verified that "the order parameter $\sigma^x$ does not relax" \cite{Jacopo}.}. Lack of relaxation of one-point functions has by now been observed in several other numerical works in $d=1$ (see e.g. \cite{KCTC,Lukin,Liu,EK}), and its search has started in $d=2$ (see \cite{Halimeh} and references therein) despite the size limitations that still affect simulations in this case.

\end{document}